\documentclass[twocolumn,prd,aps,showpacs,preprintnumbers,nofootinbib,eqsecnum]{revtex4}
\usepackage{graphics,graphicx}
\usepackage{amsmath}
\usepackage{psfrag}
\usepackage{dcolumn}
\usepackage{color}
\usepackage{fancybox}

\allowdisplaybreaks

\newcommand{\be}{\begin{equation}}
\newcommand{\ee}{\end{equation}}
\newcommand{\bea}{\begin{eqnarray}}
\newcommand{\eea}{\end{eqnarray}}

\newcommand{\vek}[1]{\boldsymbol{#1}}

\newcommand{\vL}{\vek{L}}
\newcommand{\vJ}{\vek{J}}

\newcommand{\vSone}{\vek{S}_{1}}
\newcommand{\vStwo}{\vek{S}_{2}}
\newcommand{\vS}{\vek{S}}

\newcommand{\vex}{\vek{e_X}}
\newcommand{\vey}{\vek{e_Y}}
\newcommand{\vez}{\vek{e_Z}}

\newcommand{\vi}{\vek{i}}
\newcommand{\vj}{\vek{j}}
\newcommand{\vk}{\vek{k}}

\newcommand{\vis}{\vek{i_{s}}}
\newcommand{\vjs}{\vek{j_{s}}}
\newcommand{\vks}{\vek{k_{s}}}

\newcommand{\vp}{\vek{p}}
\newcommand{\vq}{\vek{q}}
\newcommand{\vN}{\vek{N}}

\newcommand{\kSeff}{\vk \cdot \vS_\text{eff}}

\begin{document}

\title{
Gravitational waveforms from unequal-mass binaries with arbitrary spins
under leading order spin-orbit coupling
}

\author{Manuel Tessmer}
\email{M.Tessmer@uni-jena.de}
\affiliation{Theoretisch-Physikalisches Institut,
Friedrich-Schiller-Universit\"at Jena,
Max-Wien-Platz 1,
07743 Jena, Germany}

\date{\today}

\begin{abstract}
The paper generalizes the structure of gravitational waves from  
orbiting spinning binaries under leading order spin-orbit coupling, 
as given in the work by K\"onigsd\"orffer and Gopakumar
[PRD 71, 024039 (2005)] for single-spin and equal-mass binaries, to unequal-mass
binaries and arbitrary spin configurations. The orbital motion is taken 
to be quasi-circular and the fractional mass difference is assumed
to be small against one.
The emitted gravitational waveforms are given in analytic form.
\end{abstract}


\maketitle

\section{Introduction}

As already stated in many publications before, gravitational waves from
inspiralling compact binaries are the most promising sources for ground
based planned and already operating gravitational wave (GW) detectors.
To guarantee successful search for GWs, one needs to obtain promising
search templates incorporating all important physical effects that have
an influence on the form of the signal.  Ground-based detector networks
like {\em LIGO (USA)}, {\em VIRGO (France/Italy)}
and  {\em GEO 600 (Germany/UK)} have the sensitivity to be able to see the
last seconds or minutes of the binary's inspiral,
where the corrections, coming from general relativity, to the Newtonian
orbital motion get important, depending on their masses.
In order to detect GWs from inspiralling compact binaries without spin
in quasi-circular orbits, a large library of ready-to-use inspiral
templates has been put up {\cite{DIS}}. Eccentric inspiral
models without spin have also been developed {\cite{WEX_SCH, DGI2004}} and are well understood.
Recently, Yunes and collaborators obtained a formalism for frequency
domain GW filters for eccentric binaries {\cite{Yunes09_ecc_FD}}.
All of them heavily employ the post Newtonian (PN) approximation to
general relativity.

It has been shown by several authors, that for a successful detection,
effects of spin have to be included and foundations for the detection of
spins have been laid down {\cite{rapid_Kidder_Will,CutFlan94,BCV03,BCV05,Apost95}}.
During the inspiral phase, before reaching the last
stable orbit, those effects are long-term modulations of the GW signal in a comparison
with the time scale of only one orbit. They can lead to substantially
different shapes of the signal compared to those ones showing up if
the spins are neglected. 
The foundations for the motion of spins in curved spacetimes are given
in \cite{PapapetrouI}.
In {\em harmonic} coordinates, the spin dependent EOM were derived up to
next-to-leading order in the spin-orbit coupling
by Faye et al. \cite{BBFI} and  Blanchet et al. \cite{BBFII}, where
velocities have been used to characterize the orbits.
In Arnowitt-Deser-Misner coordinates {\cite{ADM}}, higher order
Hamiltonians dictating the equations of motion for orbits
and spin (from this point on referred to as EOM) have recently been
derived by Damour et al. \cite{DJS08} and Steinhoff et
al. {\cite{SSH08, SHS08}}. The spin-independent part of the binary Hamiltonian is
known to 3PN order \cite{DJS Phys. Lett. B}.\newline
The solution to ``simple precession'' of the leading order spin-orbit
interaction, which was the case for single spin or equal mass, was
discussed in {\cite{Apost95}}
and later in {\cite{KG05}}, where the GW polarizations
$h_\times$ and $h_+$
were derived as a PN accurate analytic solution for eccentric orbits.
The latter has heavily inspired this work, which will give an approach to
the more general case of unequal masses and arbitrary two-spin configurations.
\newline
The paper will be organized as follows. 
Section \ref{sec:3PN-EOM for PP & LOSO} will present the involved Hamiltonians
and the associated EOM for the binary in the center-of-mass
frame.
In section \ref{sec:Geometry}, the geometry and the coordinates
relating the generic reference frame with the orientation of the spins
and the angular momentum vector are provided and characterized by rotation
matrices.
%
%
The time derivatives of these rotation matrices will be compared by Poisson brackets
in section \ref{sec::TD and PB} and first order time derivatives
of the associated rotation angles will be obtained.
A first-order perturbative solution to the EOM for the spins is worked out in
section \ref{sec::pert_sol}.
The {\em orbital} motion will be computed, for quasi-circular orbits
(circular orbits in the precessing orbital plane), in section \ref{sec::orbital motion}.
As an application, the resulting GW polarizations, $h_\times$ and $h_+$
in the quadrupolar restriction, are given in section \ref{sec::GW polarization}.

\section{The conservative equations of motion for the spins}
\label{sec:3PN-EOM for PP & LOSO}

In this section the dynamics of spinning compact
binaries is investigated where the spin contributions are restricted to the
leading order gravitational coupling. The Hamiltonian associated therewith
reads
\begin{align}
\label{eq:dimH}
{\cal H} &
= {\cal H}_{\rm N}
+ {\cal H}_{\rm 1PN}
+ {\cal H}_{\rm 2PN}
+ {\cal H}_{\rm SO}
\,,
\end{align}
with 
${\cal H_{\rm N}}$,
${\cal H_{\rm 1PN}}$ and
${\cal H_{\rm 2PN}}$ respectively are the Newtonian, first and
second PN order contributions to the conservative
point particle dynamics {(e.g., \cite{Jaran_Schaefer}
and references therein)} and
${\cal H}_{\rm SO}$
is the leading order spin-orbit Hamiltonian {\cite{Bark_O'ConnI}}.
%
%
%

In the following computations, use will be made of the following scalings to convert
the quantities in calligraphic letters to dimensionless ones on the rhs:
 \begin{align}
 \hspace{.128\textwidth}
 {\cal H}    ~=~& {H} \, {\mu c^2} 				\,, \\
 {\cal R}    ~=~& \vek{r} \, \frac{G\,m}{c^2} \hrulefill	\,, \\
 {\cal P}    ~=~& \vek{p} \, \mu \, c				\,, \\
 {\cal S}_a  ~=~& \vek{S}_a \frac{G\, m_a}{c^2} \, (m_a\, c)	\,,
 \end{align}
where $m_a$ is the mass of the $a^{th}$ object ($a={1, 2}$), $m$ is the total mass, 
$m=m_1+m_2$, $\mu$ is the reduced mass defined as $m_1\,m_2/m$ and the symmetric
mass ratio is given by $\eta:=m_1\,m_2/m^2$. The variables
$\vek{p}$ and $\vek{r}$ are the scaled linear canonical momentum and position
vectors, respectively, and commute with the spins ${\cal S}_a$. Explicitly, the contributions
to the scaled version of
Eq. (\ref{eq:dimH}) read
 \begin{eqnarray}
\label{H_3_SO}
{H}(\vek{r}, \vek{p}, \vek{S}_{1}, \vek{S}_{2})
&
=& {H}_{\rm N}(\vek{r}, \vek{p})
+ {H}_{\rm 1PN}(\vek{r}, \vek{p})
+ {H}_{\rm 2PN}(\vek{r}, \vek{p}) \nonumber \\
&&
+~ {H}_{\rm SO}(\vek{r}, \vek{p}, \vek{S}_{1}, \vek{S}_{2})
\,,
 \end{eqnarray}
with

\begin{widetext}
\begin{subequations}
\label{H_3_Full}
\begin{align}
{H}_{\rm N}(\vek{r}, \vek{p})
&= \frac{\vek{p}^2}{2} - \frac{1}{r}
\,,
\\
{H}_{\rm 1PN}(\vek{r}, \vek{p})
&= \frac{1}{c^2} \left\{
\frac{1}{8} (3\eta-1) \left( \vek{p}^2 \right)^2
- \frac{1}{2} \left[ (3+\eta) {\vek{p}}^2 + \eta(\vek{n} \cdot
\vek{p})^2 \right] \frac{1}{r}
+ \frac{1}{2r^2} \right\}
\,,
\\
{H}_{\rm 2PN}(\vek{r}, \vek{p})
&= \frac{1}{c^4} 
\left\{
\frac{1}{16} \left( 1 - 5\eta + 5 \eta^2 \right)
\left( {\vek{p}}^2 \right)^3
+ \frac{1}{8} \left[
\left( 5 - 20 \eta - 3 \eta^2 \right)
\left({\vek{p}}^2 \right)^2
- 2 \eta^2 (\vek{n} \cdot \vek{p})^2 {\vek{p}}^2 - 3 \eta^2
(\vek{n} \cdot \vek{p})^4 \right] \frac{1}{r}
\right.
\nonumber
\\
&\quad\left.
+ \frac{1}{2} 
\left[ 
(5 + 8 \eta){\vek{p}}^2 + 3 \eta (\vek{n} \cdot \vek{p})^2 
\right] \frac{1}{r^2}
- \frac{1}{4} ( 1 + 3 \eta) \frac{1}{r^3}
\right\}
\,,
\\
{H}_{\rm SO}(\vek{r}, \vek{p}, \vek{S}_{1}, \vek{S}_{2})
&= \frac{1}{c^2 r^3} (\vek{r} {\times} \vek{p}) \cdot \vek{S}_\text{eff}
\,,
\end{align}
\end{subequations}
\end{widetext}
where $r \equiv |\vek{r}|$
and $\vek{S}_\text{eff}$ is the so-called {\em effective spin},
\begin{subequations}
\begin{align}
\label{eq:Def_eff_spin}
 \vek{S}_\text{eff}	&\equiv \delta_1 \vSone + \delta_2 \vStwo   \hfil 				\,,	\\
 \delta_1 		&\equiv \frac{\eta }{2} + \frac{3}{4} \left(1 + \sqrt{1-4 \eta }\right)		\,,	\\
 \delta_2 		&\equiv \frac{\eta }{2} + \frac{3}{4} \left(1 - \sqrt{1-4 \eta }\right)		\,.	
\end{align}
\end{subequations}
%
%
%
Considering only the spin-independent part of the Hamiltonian,
the orbital angular momentum vector is a conserved quantity.
The motion of the reduced mass $\mu$ will, without SO interactions,
take place in a plane that is perpendicular to $\vL$ and that is
invariant in time.
Adding the spin-orbit term will, in general, lead to a precession
of the orbital angular momentum.
The EOM for $\vek{L}$, defined by $\vek{L}:=\vek{r}\times\vek{p}$
and the individual spins $\vek{S_1}$ \& $\vek{S_2}$ can be deduced
from the equations
\begin{subequations}
\label{eq:d/dt_L_S1_S2_original}
\begin{flalign}
\label{eq:dLdt_with_Seff}
\frac{d \vek{L}}{dt} &
=\{ \vek{L}, {H}_\text{SO} \}
=\frac{1}{c^2 r^3} \vek{S}_\text{eff} \times \vek{L}
\,,
\\
\label{eq:dS1dt_with_L}
\frac{d \vek{S}_{1}}{dt} &
=\{ \vek{S}_{1} , {H}_\text{SO} \}
=\frac{\delta_{1} }{c^2 r^3} \vek{L} \times \vek{S}_{1}
\,,
\\
\label{eq:dS2dt_with_L}
\frac{d \vek{S}_{2}}{dt} &
=\{ \vek{S}_{2} , {H}_\text{SO} \}
=\frac{\delta_{2} }{c^2 r^3} \vek{L} \times \vek{S}_{2}
\,.
\end{flalign}
\end{subequations}
Equation (\ref{eq:dLdt_with_Seff}) describes the precession
of $\vek{L}$
w.r.t. the total angular momentum vector $\vek{J}$, defined as
{$\vJ~ \equiv ~\vL~+~\vSone~+~\vStwo$}.
The key idea in the next sections is to compute time dependent
rotation matrices for $\vL$, $\vSone$ and $\vStwo$ for a number
of rotation axes and angles that are to be introduced in the next
section.
Let us state that the magnitudes 
$L$, $S_1$ and $S_2$ of the vectors $\vL$, $\vSone$ and $\vStwo$
are conserved,
\hspace{-1cm}
\begin{subequations}
\begin{align}
\frac{d {L}^{2}}{dt}
=&~\frac{d}{dt} (\vek{L} \cdot \vek{L})
=\frac{2}{c^2 r^3} \vek{L} \cdot (\vek{S}_\text{eff} \times \vek{L} )
=0\,,
\,
\\
\frac{d S_1^2}{dt}
=&~ \frac{d}{dt} (\vek{S}_{1} \cdot \vek{S}_{1})
=\frac{2 \delta_{1} }{c^2 r^3} \vek{S}_{1} \cdot
(\vek{L} \times \vek{S}_{1})
=0\,,~
\,
\\
\frac{d S_2^2}{dt}
=&~\frac{d}{dt} (\vek{S}_{2} \cdot \vek{S}_{2})
=\frac{2 \delta_{2} }{c^2 r^3} \vek{S}_{2} \cdot
(\vek{L}\times \vek{S}_{2})
=0\,.
\,
\end{align}
\end{subequations}
Equations (\ref{eq:d/dt_L_S1_S2_original}) show that
$\dot \vL = - (\dot \vSone + \dot \vStwo)$ and, thus, the
total angular momentum vector $\vJ$
satisfies
\begin{eqnarray}
 \frac{d \vJ}{dt} = 0 
~
\text{, giving}
~
 \frac{d |\vJ|}{dt} = 0 \,.
\end{eqnarray}
The magnitudes of $\vek{S}$ and $\vek{S_\text{eff}}$ behave as follows,
\begin{subequations}
\label{eq:S_and_S_eff_mag_variation}
\begin{align}
\label{eq:Smagvariation}
\frac{d S^{2}}{dt}
&= - \frac{ 3 \sqrt{ 1 - 4 \eta } }{c^2 r^3}
\vek{L} \cdot (\vek{S}_1 \times \vek{S}_{2})
\,,
\\
\label{eq:Seff_mag_variation}
\frac{d S^2_\text{eff}}{dt}
&= - \frac{ 3 \sqrt{ 1 - 4 \eta } \, (12 + \eta)\eta}{ 4 c^2 r^3 }
\vek{L} \cdot (\vek{S}_1 \times \vek{S}_{2})
\,.
\end{align}
\end{subequations}
Notice the conservation of $S^2_\text{eff}$ in both the test-mass ($\eta
=0$) and equal-mass ($\eta = 1/4$) cases. Using above equations, we will be able to compute the
evolution equations for the rotation angles. The associated
geometry is introduced next.

\section{Geometry of the binary}
\label{sec:Geometry}
As done in \cite{KG05}, it is very useful to use a fixed
orthonormal frame ($\vex, \vey, \vez $) and to set $\vez$
along the fixed vector $\vJ$. The invariable plane
perpendicular to $\vJ$  will
then be spanned by the vectors ($\vex, \vey$). The motion
of the reduced mass will take place in the orbital plane
perpendicular to the unit vector $\vk:=\vL/L$.
For a clear understanding of the following, please take a look
at Fig. \ref{fig:sphere_xyz_ijk}.
First, the vector $\vk$ is inclined to $\vez$ by the
({\em time-dependent}) angle $\Theta$, which was also the
constant precession cone of $\vL$ around $\vJ$ for the single-spin
and equal-mass case of \cite{KG05}. As before, the orbital
plane, itself spanned by the vectors $(\vi, \vj)$, where
$\vj=\vk \times \vi$, intersects the invariable plane at
the line of nodes $\vi$, with the longitude $\Upsilon$
measured  in the invariable plane from $\vex$.

The geometry of the binary will be completed by the
spin related
coordinate system $(\vis, \vjs, \vks)$.
This frame is constructed from the system $(\vi, \vj, \vk)$
to be rotated around the axis $\vi$ to point
from the top of $\vL$ to the top of $\vJ$ with the new direction $\vks$.
In other words, this spin coordinate system is chosen in
such a way that the total spin, $\vSone + \vStwo$, has only
a $\vks$ component and $\vi_s \equiv \vi$ holds.
If $\Theta$ is known, the spins are left with an additional
freedom to rotate around $\vks$ by an angle $\phi_s$ (the
index ``s'' is a hint for positions in the spin system).
This angle is measured from $\vis$ to the projection of
$\vSone$ to the $(\vis, \vjs)$ plane, similar to $\Upsilon$'s
function in the reference frame.

There exist simple geometrical relations that will reduce
the freedom to choose rotation angles arbitrarily, as will
be shown in the next subsection.

\subsection{Geometrical issues}
As mentioned already,
in this geometry the spins and angular momenta -- being fixed in
their magnitudes -- only have three degrees of freedom: the angles
$\Theta$, $\Upsilon$ and $\phi_s$. Once $\Theta$ is determined, also
 $\alpha_\text{ks}$ (the angle between $\vL$ and $\vS$)
 is fixed and so is magnitude $S$ of $\vS =\vSone+\vStwo$
by triangular relations.
Calling $\alpha_{12}$ the angle between the spins $\vSone$ and $\vStwo$,
the following equations list the rotation angles and magnitudes
as functions of $\Theta$, where also use is made of the sin relations,
\begin{subequations}
\label{eqns:angles}
\begin{align}
\label{eq:S_tot_theta}
S(\Theta)  =~ & \sqrt{J^2-2 J L \cos \Theta + L^2} \,, \\
\nonumber \\
\label{eq:a12}
\alpha_{12}(\Theta)=~& \cos ^{-1}\left(\frac{S(\Theta)^2-S_1^2-S_2^2}{- 2 S_1 S_2}\right) \,, \\
\nonumber \\
\label{eq:ks}
\alpha_\text{ks}(\Theta)=~& \pi -\sin ^{-1}\left(\frac{J \sin (\Theta )}{\text{S}(\Theta )}\right) \,, \\
\nonumber \\
\label{eq:tlds}
\tilde{s}(\Theta)  =~& \sin ^{-1}\left(\frac{S_2 \sin \alpha_{12}(\Theta )}{\text{S}(\Theta )}\right) \,.
\end{align}
\end{subequations}

\noindent
These relations will be used extensively to simplify the angles evolution equations.
How they are incorporated and applied will be shown next.

 \begin{widetext}
\begin{center}
\begin{figure}[ct]
  \centering
%
%
%
  \includegraphics[scale=0.7, angle=90]{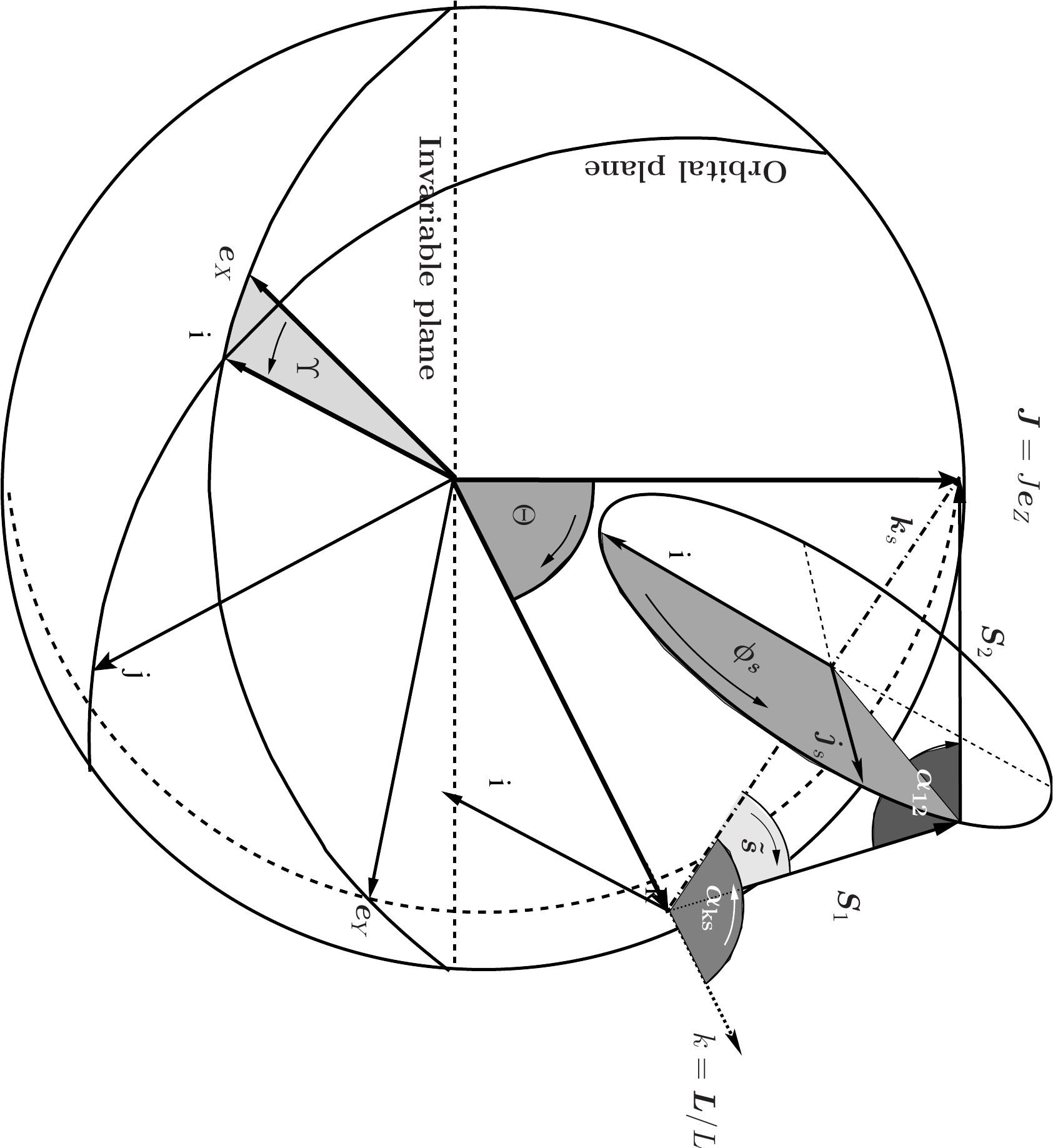}
  \caption{
	Binary geometry completed by a rotating spin coordinate
	system. The usual reference frame is ($\vex, \vey, \vez$)
	having chosen $\vez$ to be aligned with $\vJ$.
	The vectors $\vL, \vSone, \vStwo$ describe the orbital angular
	momentum and the individual spins, respectively.
	The angle $\Theta$ denotes the inclination angle of $\vL$ 
	w.r.t. $\vJ$, which is -- of course -- to be taken as a time
	dependent quantity. The orbital plane, being perpendicular
	to $\vL$ by construction, is spanned by the orthonormal
	vectors $\vj$ and $\vi$, where the latter one intersects
	the invariable plane at the angle $\Upsilon$ measured from $\vex$.
	The spin-coordinate system is constructed out of the orbital
	dreibein ($\vi, \vj, \vk$) by a rotation of $\alpha_\text{ks} $
	around $\vi$, such that the vector pointing from $\vL$ to
	$\vJ$ is the total spin $\vSone+\vStwo$. The angle $\alpha_{12}$
	is measured between $\vSone$ and $\vStwo$.
	The spin $\vSone$,
	projected into the ($\vj_s, \vi_s \equiv \vi$) plane is
	rotated by an angle $\phi_s$ from $\vi$, and $\vSone$ itself
	is moving on the circle (with variable radius) embedded in
	the figure.
  }
  \label{fig:sphere_xyz_ijk} 
\end{figure}
\end{center}
\end{widetext}

\subsection{Coordinate bases and associated transformation matrixes}

This section introduces the coordinate transformations from the reference
system to the orbital triad and the spin system. To construct the
EOM for the 3 physical angles $\Theta, \Upsilon$ and $\phi_s$, the idea is
to compare the evolution of these rotation angles - as arguments for rotation
matrices - with the Poisson brackets,
Eqs. (\ref{eq:dLdt_with_Seff}) - (\ref{eq:dS2dt_with_L}).
Let us begin with the explicit computation of the transformed coordinate
bases.
\begin{enumerate}
 \item	The orbital triad ($\vi, \vj, \vk$) can be, not surprisingly, constructed
	by only 2 rotations
	from the reference
	system. In terms of rotation matrices, we have
\begin{align}
\label{eq:ijk_explicitly}
\begin{pmatrix}
\vek{i}\\
\vek{j}\\
\vek{k}
\end{pmatrix}
=&
\begin{pmatrix}
1&0&0\\
0&\cos \Theta&\sin \Theta\\
0&-\sin \Theta&\cos \Theta
\end{pmatrix}
\begin{pmatrix}
\cos\Upsilon&\sin\Upsilon&0\\
-\sin\Upsilon&\cos\Upsilon&0\\
0&0&1
\end{pmatrix}
\\
&
\times
\begin{pmatrix}
\vek{e}_{X}\\
\vek{e}_{Y}\\
\vek{e}_{Z}
\end{pmatrix}
\,.
\nonumber
\end{align}

\item	The spin system is constructed, simply by another rotation of
	$\alpha_\text{ks} $ around the vector $\vi$, from the orbital triad,
\begin{align}
\begin{pmatrix}
\vis \\
\vjs \\
\vks 
\end{pmatrix}
&=
\begin{pmatrix}
1&0&0\\
0& \cos \alpha_\text{ks}  & -\sin \alpha_\text{ks} \\
0& \sin \alpha_\text{ks}  & ~\cos \alpha_\text{ks} 
\end{pmatrix}
\begin{pmatrix}
\vi \\
\vj \\
\vk
\end{pmatrix}
\,,
\end{align}
such that $\vis \equiv \vi$ holds. {\em Important note: the angle
$\alpha_\text{ks} $ has negative sign relative to $\Theta$. That's because
$\vks$ has to be moved ``backwards'' to point to $\vJ$! }
\end{enumerate}

Having transformed the unit vectors
with these matrices, the coordinates transform by their transposed
inverses, which are -- in case of rotations -- the matrices themselves.

Now, we have everything under control to construct the set of all
the physical vectors. I will list all of them below.
First, let me define some shorthands for rotation matrices:
\begin{eqnarray}
 \left[ \Theta \right] &\equiv&~
\begin{pmatrix}
1&0&0\\
0&\cos \Theta&\sin \Theta\\
0&-\sin \Theta&\cos \Theta
\end{pmatrix} \,, \\
\left[ \Upsilon \right] &\equiv&~
 \begin{pmatrix}
 \cos\Upsilon&\sin\Upsilon&0\\
 -\sin\Upsilon&\cos\Upsilon&0\\
 0&0&1
 \end{pmatrix} \,,  \\
\left[ \alpha_\text{ks}  \right] &\equiv&~
\begin{pmatrix}
1&0&0\\
0& \cos \alpha_\text{ks}  &-\sin \alpha_\text{ks} \\
0& \sin \alpha_\text{ks}  &~\cos \alpha_\text{ks} 
\end{pmatrix} \,.
\end{eqnarray}
The orbital angular momentum $\vL$ in the reference system
(indices labeled \emph{inv}) arises
from two rotations from
the orbital triad (\emph{ot}) where it has only one component:
\begin{equation}
\label{eq:vLwithRotMat}
 \vL =
\left \{
\left[ \Theta(t) \right] \,
\left[ \Upsilon(t) \right]
\right \}^{-1}
(0, 0, L)
\,,
\end{equation}
or, in components,
\begin{equation}
  (\vL)_{i}^\text{\rm{inv}} =
 	\biggl \{~ 
  		\left[\Theta(t) \right]
  		\left[\Upsilon(t) \right]
  		~\biggr \}^{-1}_{ij} (\vL)_{j}^\text{\rm{ot}}\,.
\end{equation}
The spins, in the spin system ({\em s}), where the $\vks$ is aligned with $\vS=\vSone+\vStwo$,
have the following form,

\begin{subequations}
\begin{align}
\hspace{-0.2cm}
	\vSone	=~& S_1( \cos \phi_s \sin \tilde s ~\vis + \sin \phi_s \sin \tilde s \, \vjs 
 	+ \cos \tilde s \, \vks )\,, \\
\hspace{-0.2cm}
\vStwo	=~& S \, \vks - \vSone \,,
\\
\hspace{-0.2cm}
\vS	=~& S \, \vks \,.
\end{align}
\end{subequations}
\section{The time derivatives of $\Upsilon$, $\Theta$ and $\phi_s$ }
\label{sec::TD and PB}
To obtain an EOM for the angle $\Theta$, one possibility is to use the
time derivative of $|S|^2 = (\vS \cdot \vS)$, Eq. (\ref{eq:Smagvariation}),
to apply this, for example, in the spin system and to compare the result
with the time  derivative of Eq. (\ref{eq:S_tot_theta}) with $\Theta=\Theta(t)$.
The result is
\begin{equation}
\label{eq:dot_Theta}
\dot \Theta = -\frac{C_\text{S} S_1 S}{2 J} \, \sin \alpha_\text{ks}  \cos \phi_s \sin \tilde s \csc \Theta
\end{equation}
with 
\begin{equation}
{C_\text{S} = - \frac{ 3 \sqrt{ 1 - 4 \eta } }{c^2 r^3}} \,.
\end{equation}
The same result will be obtained by computing the time
derivative of the orbital angular momentum $\vL$ in the
invariable system. Therefore,  take Eq. (\ref{eq:vLwithRotMat}),
compute its
time derivative and finally  compare the result with
(\ref{eq:dLdt_with_Seff}). Because
the angular velocities appear in relatively simple relations,
it is easy to extract them from the $\vex$ and $\vey$ entry.
The results are 
\begin{widetext}
\begin{eqnarray}
\label{eq:dot_Upsilon}
 \dot{\Upsilon} &=& - ~ {C_L} \csc \Theta \, \left[S_1 (\delta_2-\delta_1) \cos \alpha_\text{ks}
 \sin \phi_s  \sin \tilde s + \sin  \alpha_\text{ks} (S_1 (\delta_1-\delta_2) \cos \tilde s + {S}
   \delta_2)
\right] \,,	\\
\label{eq:dot_Theta_II}
\dot{\Theta}   &=& C_L \, S_1 (\delta_1-\delta_2) \cos \phi_s \, \sin \tilde s
\,,
\end{eqnarray}
\end{widetext}
with { $C_L := (c^2 \, r^3)^{-1}$ }. The functional dependencies
of $\alpha_\text{ks}$, $\tilde s$ and $S$ on $\Theta$ are implicated.
Inserting the geometrical relations, Eqs. (\ref{eqns:angles}), it turns
out that Eqs. (\ref{eq:dot_Theta}) and (\ref{eq:dot_Theta_II})
are equivalent. Also, the allegedly worrying  asymmetric
appearance of the quantity $S_1$ can be studiously avoided by replacing
$\tilde s$ by its function of $\Theta$.
\footnote
	{The angular velocities, (\ref{eq:dot_Upsilon}) and (\ref{eq:dot_Theta_II}),
	are in complete agreement with Eqs. (5.11a) and (5.11b) of \cite{DamSch_HORPA}.}

%
%

Also note that, if the relations $\eta=1/4$ or $S_i=0~ (i=1 \text{ or } 2)$ are inserted
in Eq. (\ref{eq:dot_Upsilon}), one recovers Eq. (4.32) of \cite{KG05}. \newline
Now, let us turn to the last quantity to be determined, the angle $\phi_s$. The
geometry offers various possibilities to calculate the time derivative of
this angle.
The easy way is to compute $\vSone$ in the invariable system. In components,
we have 
\begin{equation}
\label{eq:S1_in_INV}
  (\vSone)_{i}^\text{\rm {inv}} =
 	\biggl \{ \left[ \alpha_\text{ks} (t) \right]
  		\left[\Theta(t) \right]
  		\left[\Upsilon(t) \right]
  		  \biggr \}^{-1}_{ij} (\vSone)_{j}^\text{s} \,.
\end{equation}
The time derivative of (\ref{eq:S1_in_INV}) might be compared with Eq. (\ref{eq:dS1dt_with_L}).
The result will be given in terms of the angles already determined:
since we already know $\dot \Theta$ and $\dot \Upsilon$ on the one hand
and $\alpha_\text{ks}$ as a function of $\Theta$ on the other, we have
the expression under full control.

The other way is to take the Leibniz product rule for $\vS$, namely
$\partial_t{\vS} = \partial_t{S^i} \, \vek{e}_i + S^i \, \partial_t{\vek{e}_i}$
with $\vek{e}_i = (\vis,\vjs,\vks)$. We we already know that $\vis \equiv \vi$, $\vks || \vS$
and $\vjs \equiv \vis \times \vks$, whose time
derivatives are already known. Both considerations result in 
\begin{subequations}
\label{eq::dot_phis1}
\begin{align}
  \dot{\phi}_s~=	&~
		  {C_\Theta} \, (\dot \Theta
	~	- \dot \alpha_{\text{ks}})
	~	+ {C_{\tilde s}} \, \dot {\tilde s}
	~	+ \Omega_0	\,,
\\
 {C_\Theta}	~=&~  \tan \phi_s \,	\cot ({\alpha_\text{ks}}-\Theta )	+\sec \phi  \cot \tilde s \,,\\
 {C_{\tilde s}}	~=&~ -\sec \phi_s \,	\cot ({\alpha_\text{ks}}-\Theta )	-\tan \phi  \cot \tilde s \,, \\
 \Omega_0
		~=&~ -{C_1}\, L\, \sin \Theta \, \csc ({\alpha_\text{ks}}-\Theta)\,.
\end{align}
\end{subequations}

\noindent
For the case of equal masses ($\delta_1=\delta_2=\delta=7/8$), one obtains for $\dot \phi_s$,
$\dot \Upsilon$ and $\dot \Theta$ a very simple system of EOM,
\begin{subequations}
\begin{align}
\label{eq::angular_vel_Theta}
\dot \Theta  = & ~0\,,\\
\label{eq::angular_vel_Upsilon}
\dot \Upsilon =& ~\frac{7 J}{8 c^2 r^3}\,, \\
%
\label{eq::angular_vel_phis}
\dot \phi_s  =& ~-\frac{L \delta \sin \Theta }{c^2 r^3}
		\,
		 { \csc \left\{\sin ^{-1}\left(	\frac{J \sin \Theta }
							{S(\Theta)}
					\right)+\Theta \right\}}
\,,
\end{align}
\end{subequations}
which can be integrated immediately, giving
\begin{subequations}
\label{eq::Sol_Angles_NonPert}
\begin{align}
 \label{eq::Sol_Theta_eqm}
\Theta(t)   =& ~\Theta_{0}\,,\\
\label{eq::Sol_Upsilon_eqm}
\Upsilon(t) =& ~\Omega_\Upsilon \,t + \Upsilon_{0}\,, \\
\label{eq::Sol_Phis_eqm}
\phi_{s} (t)=& ~\Omega_\phi \,t
		+ {\phi_s}_{0}
\,,
\end{align}
\end{subequations}
with the angular velocities
\begin{subequations}
\label{eq::ang_vel}
	\begin{flalign}
\hspace{-0.2cm}
\Omega_\Upsilon \equiv &\frac{7 J}{8 c^2 r^3}\,, \hrulefill \\
\hspace{-0.2cm}
\Omega_\phi	\equiv &-\frac{L \delta \sin \Theta_0 }{c^2 r^3}
		\,
		 { \csc \left\{\sin ^{-1}\left[	\frac{J \sin \Theta_0 }
							{S(\Theta_0)}
					\right]+\Theta_0 \right\}} \,.
	\end{flalign}
\end{subequations}
Summarizing the EOM for the coordinate transformation angles,
Eqs. (\ref{eq:dot_Upsilon}), (\ref{eq:dot_Theta_II}) and (\ref{eq::dot_phis1}),
this system of EOM can be written in a compact manner. Calling the vector of constants,
${\vek {C}}=\left\{ E, S_1, S_2, L, \kSeff \right\}$ -- where $E$ and $L$ are 
related in the case of quasi-circular orbits -- and the vector of dynamic variables, associated
with spins and angular momentum,
${\vek {X}} = \left\{ \Theta, \Upsilon, \phi_s \right\}$,
we may write
\begin{align}
\frac{d}{dt}\, {\vek {X}}
=
\vek{Y}_{\vek{C}} (\vek {X})\,.
\end{align}
A perturbative solution will be given in the next section.

\section{First order perturbative solution to the EOM for the non-equal mass case}
\label{sec::pert_sol}
The EOM  for ($\Theta, \Upsilon, \phi_s$) can also be solved by a simple
reduction scheme.  We assume that the deviation from the equal-mass case
is small compared to unity,
 \begin{equation}
  \frac{\delta_1 - \delta_2}{\delta_1 + \delta_2} \ll 1  \,.
 \end{equation}
Then, having the equal-mass case under full analytic control, we can
construct a perturbative solution to the non-equal mass case.
The proceeding is as follows: Imagine a system of EOM for a number $N$
of dependent variables $\vek{X}$:
\begin{equation}
\label{eq::standard_EOM}
 \dot{ \vek{X}} = \vek{Y}(\vek{X}) \,.
\end{equation}
The time domain solution to this system is denoted by the superscript ``$0$'', viz
\begin{equation}
 \vek{X}(t) = \vek{X}^{(0)}(t)\,.
\end{equation}
Let us assume that the EOM, Eq. (\ref{eq::standard_EOM}), are perturbed by
some terms of the order $\epsilon$ ($\epsilon$ is a dimensionless ordering parameter),
\begin{equation}
\label{eq::pert_EOM}
  \dot{ \vek{X}} = \vek{Y}(\vek{X}) + \epsilon \, \vek{P}(\vek{X})\,.
\end{equation}
The solution at the first order in $\epsilon$ can be obtained by adding a small perturbed quantity to be determined
to the solution of the homogeneous equation,
\begin{equation}
   X_i^{(1)}(t) = X_i^{(0)}(t)  + \epsilon \, S_i(t)\,.
\end{equation}
Inserting this into Eq. (\ref{eq::pert_EOM}), one obtains
\begin{eqnarray}
 \dot X_i^{(1)} 	&=& \dot X_i^{(0)} + \epsilon \, \dot{S}_i\nonumber \\
		&=&Y_i (X_j ^{(0)} + \epsilon\, S_j)  + \epsilon \, P_i (X_j^{(0)} + \epsilon\, S_j) \nonumber \\
		&=&Y_i (X_j^{(0)}) + \epsilon \, \sum_{j=1}^{N} \frac{\partial Y_i}{\partial X_j} \, S_j \nonumber \\ 
		& & + \epsilon \, P_i(X_j^{(0)})	+{\cal{O}}(\epsilon^2)
\end{eqnarray}
Comparing the coefficients of the two orders of $\epsilon$ gives
\begin{eqnarray}
\label{eq::coeff_eps_0}
 0: &\dot X_i^{(0)} =& Y_i(X_j^{(0)})\,, \\
\label{eq::coeff_eps_1}
 1: &\dot S_i =& \sum_{j=1}^{N} \frac{\partial Y_i}{\partial X_j} \, S_j + P_i(X_j^{(0)})\,.
\end{eqnarray}
The first equation is solved via definition, and what remains is the second, having
inserted the unperturbed solution in the perturbing function $P$. For our purposes, $N=3$
with $\vek{X}=\{\Upsilon, \Theta, \phi_s\}$ is a small number of EOMs, but
complicated functional dependencies are included.
The matrix appearing in Eq. (\ref{eq::coeff_eps_1}) does not mean a problem to us,
because fortunately, the only dependency of the sources is on $\Theta$.

For our computation, we need to divide the EOM into a non-perturbative and a perturbative part.
In the following, we use the definitions
\begin{eqnarray}
 \chi_1 &=& \frac{\delta_1 + \delta_2}{2}\,, \\
\nonumber \\
 \chi_2 &=& \frac{\delta_1 - \delta_2}{2}\,.
\end{eqnarray}
Rewriting the EOM for the angles in terms of $\chi_1$ and $\chi_2$,
labeling all $\chi_2$ contributions with the order parameter $\epsilon$ as well as inserting
the non-perturbative solution,
Eqs. {(\ref{eq::Sol_Angles_NonPert})} to these terms, one obtains

\begin{widetext}
\begin{subequations}
\begin{align}
\dot \Theta^{(1)} = \epsilon \dot S_\Theta =&~ \epsilon\, C_L S_1 2\, \chi_2 \cos (t\,\Omega_\phi+\phi_0) \sin \tilde s(\Theta_0)
 \,, \\
\dot \Upsilon^{(0)}
+ \epsilon \dot S_\Upsilon
 =&~ 
\underbrace {C_L S(\Theta) \chi_1 \sin \alpha_\text{ks}(\Theta) \csc \Theta}
	    _{\equiv C_L J \chi_1 =\text{const.}} +
\epsilon \,
\bigl [ C_L \chi_2 \csc \Theta _0\ \bigl ( 2 S_1 
\cos \alpha_\text{ks}\left(\Theta_0 \right) 
\sin {\tilde s}\left(\Theta _0\right)
\sin \left(t \Omega _{\phi } + \phi_0\right)
\nonumber \\
&~ 
-\sin \alpha_\text{ks} \left(\Theta _0\right) \left(S(\Theta_0)-2 S_1 
 \cos \tilde s(\Theta_0) \right) \bigr ) \bigr ] \,,
\\
\dot \phi_s^{(0)}
+ \epsilon \dot S_\phi =&~ 
-C_1 L \sin
   \Theta \csc \alpha_\text{ks}(\Theta)
+ \epsilon \, \biggl [
	{C_{\tilde s}}(\Theta, \phi_s) \,\,\frac{\partial \tilde s}{\partial \Theta}
		  +
	{C_{\Theta}}(\Theta, \phi_s)
		 \, \left( 1 - \frac{\partial \alpha_\text{ks}}{\partial \Theta} \right)
	      \biggr ]
~\vline~_{^{\Theta = \Theta_0} _{\phi_s=t \, \Omega_\phi + {\phi_s}_0}} 
\, \dot \Theta \,, \\
=&~ -(\chi_1 + \epsilon\,\chi_2) 
\frac{L}{c^2 r^3} \sin
   \Theta \csc \alpha_\text{ks}(\Theta)
+ \epsilon \, \biggl [
	{C_{\tilde s}}(\Theta, \phi_s) \, \frac{\partial \tilde s}{\partial \Theta}
		  +
	{C_{\Theta}}(\Theta, \phi_s)
		 \, \left( 1 - \frac{\partial \alpha_\text{ks}}{\partial \Theta} \right)
	      \biggr ]
~\vline~_{^{\Theta = \Theta_0} _{\phi_s=t \, \Omega_\phi + {\phi_s}_0}} 
\, \dot \Theta
\,.
\end{align}
\end{subequations}
The parameter $\epsilon$ simply counts the order of the perturbative contribution and is later set to one.
The first term for $\Upsilon$ is constant and thus does not have to be expanded in powers of $\epsilon$, but the
associated first term for $\phi_s$ does, such that the perturbative solution for $\Theta$ has to be
included.
Taylor expanding this term, removing all contributions to the unperturbed problem, what remains is
a system of EOM for $S_\Theta, S_\Upsilon, S_\phi$ that can be simply integrated, because as soon
as $S_\Theta(t)$ is known, all the other contributions are straightforwardly evaluated.
Requiring that the perturbing solutions vanish at $t=0$, the solutions are simply given by
\begin{subequations}
\begin{align}
S_\Theta(t)	&= \int_{0}^{t} \dot S_\Theta   dt \,, \\
S_\Upsilon(t)	&= \int_{0}^{t} \dot S_\Upsilon dt \,, \\
S_\phi(t)	&= \int_{0}^{t} \dot S_\phi     dt \,,
\end{align}
\end{subequations}
and explicitly read
\begin{subequations}
\begin{align}
\label{eq::solut_pert}
 S_{\Theta}(t)=&~
-\frac{ C_L S_1 S_2 2 \chi_2}
{S_{(0)} \Omega _{\phi }}
\,
 {\sin \alpha_{12}}_{(0)}
 \left(\sin {\phi_s}_0 - \sin \left(t \Omega _{\phi } + {\phi_s}_0 \right)\right)
 \,,\\
S_\Upsilon(t)=&~
\frac{{C_L} {\chi_2} \csc \Theta _0 }
{\Omega _{\phi }}
\biggl [2 S_1
 \cos {\alpha_\text{ks}}_{(0)} \sin \tilde s_{(0)}
 \left(\cos {\phi_s}_0
 - \cos \left(t \Omega_{\phi } + {\phi_s}_0 \right)\right)
\nonumber \\
&~
-t \Omega _{\phi } \sin {\alpha_\text{ks}}_{(0)}
 \left(S_{(0)}
 -2 S_1 \cos \tilde s_{(0)}\right)
\biggr ]
 \,, \\
S_\phi(t) =&~
C_\text{stat} \, t + C_0(t) + C_{\tilde s}(t) + C_\Theta (t) + C_{\alpha} (t) \,,
\end{align}
\end{subequations}
with the shorthands
\begin{subequations}
\begin{align}
 C_\text{stat}		=&~ \frac{ \chi _2 \Omega _{\phi }}{\chi _1}
 \,, \\
\nonumber \\
 C_0 (t) 		=&~ \frac{{C_L} J S_1 S_2 \chi _2 \sin \Theta_0
			\sin {\alpha_{12}}_{(0)} \left(t \Omega _{\phi }
			 \sin {\phi_s}_0
		 + \cos \left(t \Omega _{\phi } + {\phi_s}_0 \right)
		 - \cos \phi _0 \right) \left(-2 c^2 J r^3
\Omega^*
	-2 \chi _1 S_{(0)}^2\right)}{\chi _1 S_{(0)}^3 \Omega _{\phi } \sqrt{S_{(0)}^2-J^2 \sin
   ^2 \Theta_0}}
 \,, \\
\nonumber\\
 C_{\tilde s}(t)	=&~ \frac{-2 {C_L} J \chi _2
 \left(
S_{(0)}^2 \cot {\alpha_{12}}_{(0)}-S_1 S_2 \sin {\alpha_{12}}_{(0)}
\right)
}
{\chi _1 S_{(0)}^2 \Omega _{\phi } \sqrt{S_{(0)}^2
 - S_2^2 \sin ^2{\alpha_{12}}_{(0)}}}
\times
\nonumber\\
&~{ \left(c^2 r^3 t \Omega _{\phi } \sin \tilde s_{(0)}
\Omega^*
+L \chi _1 \sin \Theta_0  \cos \tilde s_{(0)}
	\left(
		 \cos {\phi_s}_0
		- \cos \left(t \Omega_{\phi} + {\phi_s}_0 \right)
	\right)
\right)}
 \,, \\
\nonumber\\
 C_{\Theta}(t)		=&~
	\frac{2 S_1 t \chi _2 \cos \tilde s_{(0)}}{c^2 r^3}
+\frac{2 S_1 \chi _2 \Omega^*  \csc \Theta_0 \sin
   \tilde s_{(0)} \left(\cos {\phi_s}_0 -\cos \left(t \Omega _{\phi } + {\phi_s}_0 \right)\right)}{L \chi_1 \Omega
   _{\phi }}
\,,
\nonumber\\
C_{\alpha}(t)		=&~
 C_\Theta (t)
 \,
\frac{J \left(S_{(0)}^2 \cos  \Theta_0 - J L \sin ^2 \Theta_0 \right)}{S_{(0)}^2
   \sqrt{S_{(0)}^2-J^2 \sin ^2 \Theta_0 }}
\,,
\end{align}
\end{subequations}
the initial values of the functions (\ref{eq:S_tot_theta}) - (\ref{eq:tlds})
\begin{subequations}
\begin{align}
  S_{(0)}			&\equiv S(\Theta_{0})		\,, \\
  {\alpha_\text{ks}}_{(0)}	&\equiv \alpha_\text{ks}(\Theta_0)	\,, \\
  {\alpha_\text{12}}_{(0)}	&\equiv \alpha_\text{12}(\Theta_0)	\,, \\
  {\tilde s}_{(0)}		&\equiv \tilde s (\Theta_0)		\,, \\
\end{align}
and the definition
\begin{align}
  \Omega^*			&\equiv \Omega_{\phi} \, \sqrt{1-\frac{L^2 \chi _1^2 \sin ^2 \Theta_0}{c^4 r^6 \Omega _{\phi }^2}} \,.
\end{align}
\end{subequations}

\end{widetext}

\section{The orbital motion}
\label{sec::orbital motion}
\noindent
The motion of the spins is only half of the physical content of the spin-orbit
dynamics. Once we fully have the motion of all the spin-related angles under control, we
might turn to the orbital dynamics, {\em i.e.} the motion of the reduced mass in the
orbital plane.
It will turn out that employing coordinate transformations will be very helpful here, too.

The aim is to solve the {\em orbital} EOM to the full Hamiltonian,
\begin{equation}
\label{eq::H_12_SO}
 {H}		= H_\text{N}
		+H_\text{1PN}+H_\text{2PN}
		+H_\text{SO}\,.
\end{equation}
At this point, we can do a useful simplification.
As long as we incorporate only leading order spin dynamics, only Newtonian
point particle and spin dependent contributions will mix at the end, higher
order PN terms coupling with the spins will be neglected consequently.
For the computation of the spin dependent part of the orbital phase, therefore,
we only have to take ${H}_{\rm N, SO}= H_{\rm N} + H_{\rm SO}$ and add the 1PN and
2PN (spinless) terms for the point particle afterwards.
\begin{eqnarray}
 H &=& {H}_{\rm N, SO} + H_{\rm 1PN} + H_{\rm 2PN}\,, \\
 \dot \varphi
  &=& \dot \varphi_{\rm N,SO} + \dot \varphi_{\rm 1PN} + \dot \varphi_{\rm 2PN} \,.
\end{eqnarray}

\begin{center}
\begin{figure}[!ht]
  \centering
  \includegraphics[scale=0.45, angle=90]{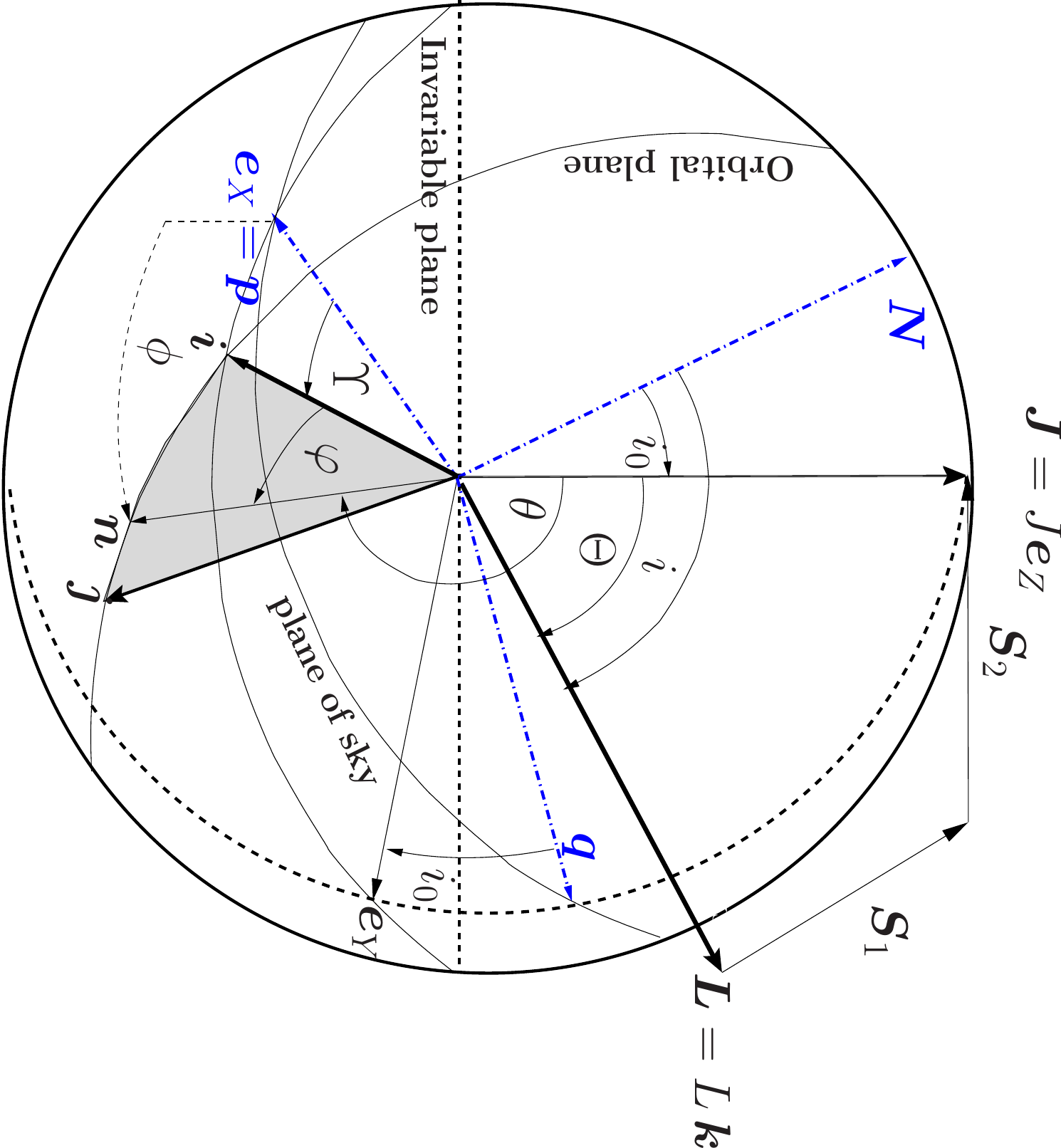}
  \caption{
	The geometry of the binary, having added the observer related
	frame $(\vek{p},\vek{q},\vek{N})$ (in dashed and dotted lines)
	with $\vek{N}$ as the line--of--sight vector, after removing the
	angles in the spin frame. The line--of--sight vector is chosen
	to lie in the $\vek{e}_{Y}$--$\vek{e}_{Z}$--plane,
	and measures an angle $i_0$ (associated with the rotation around
	$\vex$) from $\vez$, such that $\vek{p} = \vek{e}_{X}$, and this
	is the point where the orbital plane meets the plane of the sky.
	Because of this rotation, the angle $i_0$ is also found between
	the vector $\vq$, itself positioned in $(\vey, \vez)$, too, and $\vey$.
	The grey area in the graphics completely lies in the orbital plane,
	spanned by $(\vi, \vj)$ and $\varphi$ measures the angle between
	the separation vector $\vek{r}$ and $\vi$.
	The polarization vectors $\vek{p}$ and $\vek{q}$ span the plane of
	the sky. The inclination of this plane with respect to the orbital
	plane is the \emph{orbital} inclination $i$.
	The inclination of the orbital plane with respect to the invariable
	plane is denoted by $\Theta$.
	Please note that $\vL$ does {\em not} lie on the unit sphere, only
	$\vk$ does!
  }
  \label{fig:sphere_xyz_pqN} 
\end{figure}
\end{center}

The Newtonian and spin orbit part of eq. (\ref{eq::H_12_SO}) reads
\begin{equation}
\label{eq:HN_SO}
 {H}_{\rm N, SO}
		= \frac{\vek{p}^2}{2} - \frac{1}{r}
		+ \frac{1}{c^2 r^3} (\vek{r} {\times} \vek{p}) \cdot \vek{S}_\text{eff}
\,.
\end{equation}
and can be handled with the method described in \cite{KG05}.
The aim there was to introduce advantageous spherical coordinates,
$(r, \theta, \phi),$ with their associated ONS $(\vek{n}, \vek{e_\theta}, \vek{e_\phi} )$
with $\vez \cdot \vek{n}=\cos{\theta}$, $\vek n \cdot \vex = \cos \phi \, \sin \theta$, as
can be seen in Fig. (\ref{fig:sphere_xyz_pqN}).
First, we define the normalized relative separation vector according to
\begin{align}
\label{eq:vec_n_in_r_theta_phi}
\vek{n} = 
\sin\theta \cos\phi \, \vek{e}_{X} 
+ \sin\theta \sin\phi \, \vek{e}_{Y} 
+ \cos\theta \vek{e}_{Z}
\,.
\end{align}
The time derivative of $\vek{r}$, the linear momentum $\vek{p}$, its decomposition
in radial components and the corresponding orthogonal ones can be written as
%
\begin{subequations}
\begin{align}
\label{eq:vec_r_in_r_theta_phi}
\vek{r}
&= r \vek{n}
\,,
\\
\dot{\vek{r}} 
&= \dot{r} \vek{n} 
+ r \dot{\theta} \, \vek{e}_{\theta} 
+ r \sin\theta \dot{\phi} \, \vek{e}_{\phi}
\,,\\
\vek{p} &=
p_r \vek{n} + p_\theta \vek{e}_\theta + p_\phi \vek{e}_\phi \,,
\\
\vek{p}^2
&= p_r^2 + p_\theta^2 + p_\phi^2
= (\vek{n} \cdot \vek{p})^2 + (\vek{n} \times \vek{p})^2
\nonumber
\\
&= p_r^2 + \frac{L^2}{r^2} \,.
\end{align}
\end{subequations}

\noindent
Inserting $p^2$ into Eq. (\ref{eq:HN_SO}), computing $p_\phi = \vek{p} \cdot \vek{e_\phi}$
and using the orthogonality relation of the used triad, one obtains
\begin{subequations}
\label{eq:comp_of_p_with_cons_quan}
\begin{align}
\label{eq:comp_pr_of_p}
p_r^2 &= 2 E + \frac{2}{r} - \frac{L^2}{r^2}
- \frac{2 ( \vek{L} \cdot \vek{S}_\text{eff}) }{c^2 r^3}
\,,
\\
\label{eq:comp_pphi_of_p}
p_\phi &= \frac{ L_{z} }{r \sin\theta}
\,,
\\
\label{eq:comp_ptheta_of_p}
p_\theta^2 &= \frac{L^2}{r^2}-p_\phi^2
= \frac{1}{r^2} \left(L^2 - \frac{L_{z}^2}{\sin^2 \theta} \right)
\,,
\end{align}
\end{subequations}
In \cite{KG05}, it was possible to reduce these equations by some
algebraic relations and the fact that the angle $\Theta$
was constant in time - here, it is more complicated. It is
still allowed to express $L_z$, the projection of $\vL$ onto $\vez$,
in Eq. (\ref{eq:comp_pphi_of_p})
and (\ref{eq:comp_ptheta_of_p}) over $\Theta$ with the help of
\begin{subequations}
\label{eq:p_phi_with_Theta_AND_p_theta2_with_Theta}
\begin{align}
\label{eq:p_phi_with_Theta}
p_\phi 
&= \frac{L}{r} \frac{\cos\Theta}{\sin\theta}
\,,
\\
\label{eq:p_theta2_with_Theta}
p_{\theta}^2 
&= \frac{L^2}{r^2} \left(1 - \frac{\cos^2 \Theta}{\sin^2 \theta} \right)
\,.
\end{align}
\end{subequations}
Above equations are, for our purposes, the most simplified versions of the $p$ components
and will enter in the dynamics of the angle $\varphi$ in their current form.

Our aim is now to connect the coordinate velocities, namely $\dot r, \dot \phi$
and $\dot \theta$, to conserved quantities associated with the Hamiltonian
of Eq. (\ref{eq:HN_SO}).
Computation of the velocity in spherical coordinates, Eq. (\ref{eq:vec_n_in_r_theta_phi}),
gives following formulae using Hamilton's EOM,
$\dot{\vek{r}} = \partial {H}_\text{NSO}/\partial \vek{p}$, 
$\vek{n} \times \vek{e}_\theta = \vek{e}_\phi$ and
$\vek{n} \times \vek{e}_\phi = - \vek{e}_\theta$.
\begin{subequations}
\label{eq:eom_newton_with_so}
\begin{align}
\label{eq:eom_newton_with_so_1}
\dot{r} 
&= \vek{n} \cdot \dot{\vek{r}}
= p_r
\,,
\\
\label{eq:eom_newton_with_so_2}
r \dot{\theta} 
&= \vek{e}_\theta \cdot \dot{\vek{r}}
= p_\theta + \frac{ \vek{e}_\phi \cdot \vek{S}_\text{eff} }{c^2 r^2} 
\,,
\\
\label{eq:eom_newton_with_so_3}
r \sin \theta  \dot{\phi} 
&=  \vek{e}_\phi \cdot \dot{\vek{r}} 
= p_\phi - \frac{ \vek{e}_\theta \cdot \vek{S}_\text{eff} }{c^2 r^2} 
\,.
\end{align}
\end{subequations}
Of course, in the case of quasi-circular motion,
$\dot r = 0 = p_r$ holds for all times.
Remembering the geometry of Fig. \ref{fig:sphere_xyz_pqN},
we recall that $\vek r$ is lying in the plane orthogonal
to $\vL$, which itself is spanned by the vectors $\vi$
and $\vj$. Calling $\varphi$ (the orbital phase) the measure for the angular
distance from $\vi$, we can write
\begin{align}
\label{eq:vector_r_in_ijk_frame}
\vek{r} &=
r \cos\varphi \, \vek{i}
+ r \sin\varphi \, \vek{j}
\,.
\end{align}
The comparison of $\vek{r}$, given by Eqs.~\eqref{eq:vec_r_in_r_theta_phi} and 
\eqref{eq:vec_n_in_r_theta_phi}, with
the one in the new angular variables, Eq.~\eqref{eq:vector_r_in_ijk_frame} with Eqs.~\eqref{eq:ijk_explicitly}, implies the transformation
\begin{equation}
\label{eq:transfo_angular_coordinates}
	(\theta, \phi) \rightarrow (\Upsilon, \varphi):
	\begin{cases}
		\cos \theta = \sin \varphi \sin \Theta
		\\
		\sin(\phi-\Upsilon) \sin \theta = \sin \varphi \cos \Theta
		\\
		\cos(\phi-\Upsilon) \sin \theta = \cos \varphi  \,.
	\end{cases}
\end{equation}
Time derivation of the first equation will give an expression for $\dot \theta$,
which can be simplified using the third one. The final expression is
\begin{equation}
 \dot \theta =	- \sin \Delta \, \dot \Theta
		- \sqrt{1- \frac{\cos ^2 \Theta} {\sin^2 \theta} } \, \dot \varphi
\end{equation}
\noindent
with $\Delta \equiv \phi-\Upsilon$.
Setting $\Theta$ constant, one naturally recovers Eq. (4.28a) of \cite{KG05}.
Using this equation to eliminate $\dot \theta$ in (\ref{eq:eom_newton_with_so_2})
and after substition {$\pm p_{\theta}$} from (\ref{eq:p_theta2_with_Theta}), one obtains a
solution for $\dot \varphi$ and $\dot \Upsilon$
\begin{equation}
\label{eq:dot_varphi_with_L}
 \dot \varphi =    {\mp} \frac{L}{r^2}
		 - \frac{\tilde S_{ \phi}}{\sqrt{1-\frac{\cos^2 \Theta}{\sin^2 \theta}}} \frac{1}{c^2 \, r^3}
		 - \frac{\sin \Delta}{\sqrt{1-\frac{\cos^2 \Theta}{\sin^2 \theta}}} \, \dot \Theta \,,
\end{equation}
  where $\tilde S_{ \phi}$ is a shorthand for $\vek{S}_\text{eff} \cdot \vek{e_\phi}$.
{
	{
	The ambiguity of the sign in the first term can be removed if one takes
	the rotation sense of the reduced mass, or equivalently, the direction of
	$\vL$ into account. Having (initially) the vector $\vL$ in the northern
	hemisphere, one should choose ``$+$'' in above equation. This condition
	then holds anytime
	as long
	as $S_1 + S_2 < \sqrt{L^2 + J^2}$.
}}

The quantity $L/r^2$ represents only the Newtonian point particle contribution.
%
%
To express $r$ and $L$ in Eq. (\ref{eq:dot_varphi_with_L}) in terms of $E$, one only
needs Newtonian order,
\begin{eqnarray}
\label{eq:Newton+SO_radial_dist}
r &=& (-2E)^{-1}		\,,\\
L &=& (-2E)^{-1/2}	\,.
\end{eqnarray}
Summarizing the evolution for $\dot \varphi$, one can separate it into a pure point particle (PP)
and the spin orbit part (SO),
\begin{equation}
\dot \varphi = \dot \varphi_\text{PP} + \dot \varphi_\text{SO} \,.
\end{equation}
The full 2PN expression for $\dot \varphi_\text{PP}$ can be extracted from
Eqs.(5.6c), (5.6d) and (5.6k) of \cite{KG05} without spin dependent terms,
\begin{widetext}
\begin{subequations}
 \begin{align}
n &= (- 2 E)^{3/2} \bigg\{ 1 + \frac{ (-2 E) }{ 8 c^2 }  
\left( - 15 + \eta \right)
+
\frac{ (-2 E)^2 }{128 c^4 } 
\biggl[555 + 30 \eta + 11 \eta^2 
- \frac{ 192 }{ \sqrt{- 2 E L^2} } ( 5 - 2 \eta )
\biggr ]
\bigg\}
\,,
\\
k
&= \frac{3}{c^2 L^2}
\biggl\{
1
 + \frac{ (-2 E) }{4 c^2}
 \left( - 5 + 2 \eta + \frac{ 35 - 10 \eta }{ -2 E L^2 } \right)
 \biggr\}
 \,,
 \end{align}
\end{subequations}
setting $e_t = 0$ in
\begin{align}
 e_t^2
=&~ 1+ 2 E L^2 + \frac{ -2 E }{ 4 c^2 }
\bigg\{ 
- 8 + 8 \eta + (17 - 7 \eta ) (-2 E L^2) - 8 \chi_\text{so} \cos\alpha \frac{S}{L}
\bigg\} 
\nonumber
\\
&~+ \frac{ (-2 E)^2 }{ 8 c^4 } \bigg\{ 8 + 4 \eta + 20 \eta^2
- (- 2 E L^2) ( 112 - 47 \eta + 16 \eta^2 )
+ 24 \sqrt{- 2 E L^2} (5 - 2 \eta) 
\nonumber
\\
 \quad    
&~+ \frac{4}{ (-2 E L^2) } \left( 17 - 11 \eta \right)
- \frac{24}{ \sqrt{ - 2 E L^2 } } \left( 5 - 2 \eta \right)
\bigg\}
\end{align}
to eliminate $L$ and using $\dot \varphi_\text{PP} = n\, (1+k)$ \cite{DD85}, giving
\begin{align}
\label{eq:dot_varphi_PP}
\dot \varphi_{\text{PP}}
=&~ 
(-2E)^{3/2}
\biggl \{
1
+ {\frac {1}{8}}
\left( {9}+\eta \right) \frac{(-2E)}{c^2}
+ \biggl [
{\frac {891}{128}} 
-{\frac {201}{ 64}}\,\eta
+{\frac {11}{128}}\,{\eta}^{2} 
\biggr ]
\frac{{(-2E)}^{2}}{c^4}
%
\biggr \} \,,
\\
\label{eq:dot_varphi_SO}
\dot \varphi_\text{SO}
 =&~
{
		- 3 (\kSeff) {(-2 E)}^3
		+\frac{(-2E)^3 \tilde{S}_\phi}{{\sqrt{1-\frac{\cos^2 \Theta}{\sin^2 \theta}}}}
		- \frac{\sin \Delta \dot \Theta}{{\sqrt{1-\frac{\cos^2 \Theta}{\sin^2 \theta}}}}
\,,
}
\end{align}
with
\begin{align}
\tilde S_{\phi} \equiv&~ \vek{S}_{\text{eff}} \cdot \vek{e}_{\phi}\nonumber\\
 		=&~	\cos (\phi -\Upsilon )
				 [{S_1} \sin \phi_s \sin s (\delta_1-\delta_2) \cos (\Theta -{\alpha_\text{ks}})
					-\sin (\Theta -{\alpha_\text{ks}}) ({S_1} \cos \tilde s (\delta_1-\delta_2)+{S(\Theta)} \delta_2)]
\nonumber \\
&~			+\sin (\phi -\Upsilon ) {S_1} \cos \phi_s  \sin s (\delta_2-\delta_1) \,.
\end{align}
{The first term in $\dot \varphi_{SO}$, Eq. (\ref{eq:dot_varphi_SO}), comes
from spin-orbit contributions to the value of $L$, as  this is obtained from
the energy expression (\ref{H_3_SO}), see section IV of \cite{KG05}}.
The angle $\phi$ can be computed with the help of Eq. (\ref{eq:transfo_angular_coordinates})
according to
\begin{equation}
\phi=\Upsilon + \arccos \left( \cos \varphi /\sqrt{1-\sin^2 \varphi \sin^2 \Theta} \right) \,.
\end{equation}
Inserting the solutions $\Theta(t), \Upsilon(t)$ and $\phi_{s}(t)$ from sections
\ref{sec::TD and PB} and \ref{sec::pert_sol} to Eqs. (\ref{eq:dot_varphi_PP}) and (\ref{eq:dot_varphi_SO}), $\varphi$ can be obtained by numerical integration,
\begin{equation}
 \varphi(t)	= \int_{0}^{t} \dot \varphi \, \mbox{d}t +\varphi_0
		= \dot \varphi_{\text{PP}} \, t
		 + \int_{0}^{t} \dot \varphi_\text{SO}(t) \, \mbox{d}t +\varphi_0 \,.
\end{equation}
The radial separation at 2PN accuracy, after eliminating L, reads
\begin{align}
r
=&~ \frac{1}{(-2E)}
 \biggl \{
1
+ \frac{(-2E)}{4 c^2}  \biggl[\eta -7 +4 \, {(\kSeff)} \sqrt{(-2E)} \biggr]
+ \frac{(-2E)^2 }{16 c^4} \biggl[-67+\eta  (54 + \eta) \biggr] 
\biggr \}
\,.
\end{align}

\end{widetext}


\section{Gravitational waveforms}
\label{sec::GW polarization}
%
%
\noindent
The gravitational wave polarization states, $h_{+}$ and $h_{\times}$,
are usually given by 
\begin{subequations}
\label{eq:definition_hp_hx}
\begin{align}
h_{+} &= \frac{1}{2} \left(p_i p_j - q_i q_j \right) h_{ij}^\text{TT}
\,, \\
h_{\times} &= \frac{1}{2} \left(p_i q_j + p_j q_i \right) h_{ij}^\text{TT}
\,,
\end{align}
\end{subequations}
where $p_i$ and $q_i$ are the components of the vectors $\vek{p}$ and
$\vek{q}$ orthogonal to the observer's direction, respectively, and
$h_{ij}^\text{TT}$ is the transverse and traceless part of the radiation
field $h_{ij}$.
%
%
%
The leading order contribution, $h^{TT}_{ij}|_Q$, where the subscript
$Q$ denotes {\em quadrupolar} approximation, reads \cite{WillWise96}
\begin{align}
\label{eq:definition_h_newton}
h_{km}^\text{TT} \big|_{\text Q}
&= \frac{4 G \mu }{ c^4 R'} {\cal P}_{kmij}(\vek{N})
\left( v_{ij} - \frac{G M}{r} n_{ij} \right)
\,,
\end{align}
with ${\cal P}_{kmij}(\vek{N})$ as the usual transverse-traceless projection
orthogonal to the line-of-sight vector $\vek{N}$, $R'$ as the radial distance
to the binary, the shorthands $v_{ij}\equiv v_i\,v_j$ and $n_{ij}\equiv n_i\,n_j$, using
$\vek v\equiv d\vek{r}/dt$ as the velocity vector and $\vek{n}\equiv \vek{r}/r$ as the
normalized relative separation, respectively.

Using Eq. (\ref{eq:definition_h_newton}), one may express both amplitudes of
$h_\times$ and $h_+$ as
\begin{subequations}
\label{eq:h_plus_and_h_cross_in_n_and_v}
\begin{align}
h_{+} \big|_{\text Q}
&= \frac{2 G \mu}{c^4 R'}
\left[ \left( p_i p_j - q_i q_j \right)
\left( v_{ij} - \frac{G M}{r} n_{ij} \right) \right]
\nonumber
\\
&= \frac{2 G \mu}{c^4 R'}
\bigg\{ 
(\vek{p} \cdot \vek{v} )^2 - (\vek{q} \cdot \vek{v} )^2
\nonumber
\\
&\quad
-  \frac{G M}{r} \left[ ( \vek{p} \cdot \vek{n} )^2
- ( \vek{q} \cdot \vek{n} )^2 \right] 
\bigg\}
\,, \\
h_{\times} \big|_{\text Q}
&= \frac{2 G \mu}{c^4 R'}
\biggl[ \left( p_i q_j + p_j q_i \right)
\left( v_{ij} - \frac{G M}{r} n_{ij} \right) \biggr]
\nonumber \\
\nonumber \\
&= \frac{4 G \mu}{c^4 R'}
\biggl\{ (\vek{p} \cdot \vek{v}) (\vek{q} \cdot \vek{v}) - \frac{G M}{r}
(\vek{p} \cdot \vek{n}) (\vek{q} \cdot \vek{n}) \biggr\}
\,.
\end{align}
\end{subequations}
To compute the two gravitational wave polarizations, one requires an expression
for the radial separation vector $\vek{r}$ and its first
time derivative. It is efficient to give $\vek{r}$
expanded in the observer's triad $(\vp, \vq, \vN)$.
In \cite{KG05}, this was done by expressing $\vek{r}$
in $(\vex, \vey, \vez)$ first, and secondly to compute this base
from $(\vp, \vq, \vN)$ as rotated around $\vp$ with the
(constant) angle $i_0$. The result reads
%
\begin{align}
\label{eq:vector_r_in_pqN}
\vek{r} &=
r \, \bigl[
 \left\{\cos\Upsilon \cos\varphi - C_\Theta \sin\Upsilon \sin\varphi\right\}
\vek{p}
\nonumber\\
&\quad
+ \left\{
C_{i_{0}} \sin\Upsilon \cos\varphi
- \left( S_{i_{0}} S_\Theta - C_{i_{0}} C_\Theta \cos\Upsilon \right)
\sin\varphi 
\right\} \vek{q}
\nonumber\\
& \quad
+ \left\{S_{i_{0}} \sin\Upsilon \cos\varphi
+ \left( C_{i_{0}} S_\Theta + S_{i_{0}} C_\Theta \cos\Upsilon \right)
\sin\varphi 
\right\} \vek{N}
\bigr]
\,,
\end{align}
where $C_{i_{0}}$ and $S_{i_{0}}$ are shorthands for $\cos i_0$ and $\sin i_0$,
respectively.
The velocity vector $\vek{v}=d \vek{r}/dt$ is given by
\begin{widetext}
\begin{eqnarray}
\label{eq:vector_v_in_pqN}
\vek{v}  &=&
r \, \bigl[
	\bigl \{ \dot \Theta \sin \Theta  \sin \Upsilon \sin \varphi
		-\dot \Upsilon (\cos \Theta  \cos \Upsilon \sin\varphi+\sin \Upsilon \cos \varphi) \nonumber \\
	 &&
	-\dot \varphi (\cos \Theta  \sin \Upsilon \cos \varphi+\cos \Upsilon \sin \varphi)
		\bigr \} \, \vek{p}
		\nonumber \\
	&& +\bigl \{
		 \dot \Theta \sin \varphi (-(C_{i_0} \sin \Theta  \cos \Upsilon+S_{i_0} \cos \Theta))
		+C_{i_0} \dot \Upsilon (\cos \Upsilon \cos \varphi-\cos \Theta  \sin \Upsilon \sin \varphi)\nonumber \\
	&&	+\dot \varphi (\cos \varphi (C_{i_0} \cos \Theta  \cos \Upsilon-S_{i_0} \sin
   \Theta )-C_{i_0} \sin \Upsilon \sin \varphi)
	\bigr \} \, \vek{q}
		\nonumber \\
 	&& +\bigl\{
		 \dot \Theta \sin \varphi (C_{i_0} \cos \Theta -S_{i_0} \sin \Theta  \cos \Upsilon)
		+S_{i_0} \dot \Upsilon (\cos \Upsilon \cos \varphi-\cos \Theta  \sin \Upsilon \sin \varphi)\nonumber \\
	&&	+\dot \varphi (\cos \varphi (S_{i_0} \cos \Theta  \cos \Upsilon+C_{i_0} \sin
    \Theta )-S_{i_0} \sin \Upsilon \sin \varphi)
	\bigr\} \, \vek{N}
\bigr ]
\,.
\end{eqnarray}
Having inserted above equations into (\ref{eq:h_plus_and_h_cross_in_n_and_v}),
the final expressions for $h_\times$ and $h_+$ with time dependent $\Theta$
and the case of quasi-circular orbits are given by
\begin{eqnarray}
\label{eq::hx_full}
h_\times|_\text{Q}
	^{[\dot r \equiv 0]}= &&
 \frac{2 G \mu}{c^4 R} \Bigl\{
-\frac{G m}{r}
   \bigl[(\cos \Upsilon \cos \varphi-\cos \Theta \sin \Upsilon \sin \varphi) (C_{i_0} (\cos \Theta
   \cos \Upsilon \sin \varphi+\sin \Upsilon \cos \varphi)
\nonumber \\
&&
-S_{i_0} \sin \Theta \sin \varphi )\bigr]
\nonumber \\
&&
+   r^2 \bigl [-\bigl ({\dot \Theta} \sin \Theta \sin \Upsilon \sin \varphi-{\dot \Upsilon} (\cos \Theta
   \cos \Upsilon \sin \varphi+\sin \Upsilon \cos \varphi)-{\dot \varphi} (\cos \Theta \sin \Upsilon \cos
   \varphi
\nonumber\\
&&
+\cos \Upsilon \sin \varphi)\bigr) \bigr] \bigl({\dot \Theta} \sin \varphi (C_{i_0} \sin (\Theta
   ) \cos \Upsilon+S_{i_0} \cos \Theta)+C_{i_0} {\dot \Upsilon} (\cos \Theta \sin \Upsilon \sin
   \varphi
\nonumber\\
&&
-\cos \Upsilon \cos \varphi)+{\dot \varphi} (-C_{i_0} \cos \Theta \cos \Upsilon \cos (\varphi
   )+S_{i_0} \sin \Theta \cos \varphi
\nonumber\\
&&
+C_{i_0} \sin \Upsilon \sin \varphi)\bigr)
   \Bigr\}
\,,
\end{eqnarray}
\begin{eqnarray}
\label{eq::h+_full}
 {h_+}|_\text{Q}
	^{[\dot r \equiv 0]}
=&&
\frac{2 G \mu}{c^4 R} \biggl \{\frac{-G m}{r} \bigl[(\cos \Upsilon \cos \varphi-\cos \Theta \sin \Upsilon \sin \varphi)^2
-(\sin
   i_0 \sin \Theta \sin \varphi
\nonumber \\
&&
-C_{i_0} (\cos \Theta \cos \Upsilon \sin \varphi
   +\sin\Upsilon \cos \varphi))^2\bigr]
\nonumber\\
&&
-r^2 \bigl[{\dot \Theta} \sin \varphi (C_{i_0} \sin \Theta \cos
   \Upsilon+S_{i_0} \cos \Theta)+C_{i_0} {\dot \Upsilon} (\cos \Theta \sin \Upsilon \sin (\varphi
   )-\cos \Upsilon \cos \varphi)
\nonumber \\
&&
+{\dot \varphi} (-C_{i_0} \cos \Theta \cos \Upsilon \cos \varphi+\sin
   i_0 \sin \Theta \cos \varphi+C_{i_0} \sin \Upsilon \sin \varphi)\bigr]^2
\nonumber \\
&&
+r^2 \bigl[{\dot \Theta}
   (-\sin \Theta) \sin \Upsilon \sin \varphi+{\dot \Upsilon} (\cos \Theta \cos \Upsilon \sin \varphi+\sin
   \Upsilon \cos \varphi)
\nonumber \\
&&
+{\dot \varphi} (\cos \Theta \sin \Upsilon \cos \varphi+\cos \Upsilon \sin \varphi
   )\bigr]^2
\biggr\}
\,.
\end{eqnarray}
\end{widetext}

\section{Conclusions and outlook}
\noindent
In this paper, the EOM of the spins and the orbital phase for the
conservative 2PN accurate point particle were solved for the case of
quasi-circular orbits, including the leading order spin-orbit interaction. The
associated gravitational waveforms, $h_{+}$ and $h_{\times}$,
in  the  quadrupolar restriction are given in analytic form.
The spins are characterized by their constant magnitudes and 3 essential
dynamic configuration angles, whose first order time derivatives were
computed with the aid of Poisson brackets, and appear to decouple from the
orbital phase.
Although these equations are quite complicated and have to be integrated
numerically in general, they reduce to quite  simple ones in the case of
equal masses and are then able to be solved exactly.

For small deviations from equal masses, a simple perturbative reduction
scheme for the EOM can be employed. The associated first order corrections
to the unperturbed equal-mass solution have been derived.
The reliability of this solution naturally depends on the precision of measurement.
The corrections are of the same PN order as the unperturbed
solutions, multiplied by a factor of
$F=(\delta_1 - \delta_2)/(\delta_1 + \delta_2)$.
If we set $m_2~=~m_1\,(1~+~\alpha)$, we obtain following representative
pairs $(\alpha, F(\alpha))$:
(0.1,~0.04),
(0.2,~0.08),
(0.5,~0.17),
(1.0,~0.29),
to give an estimate of the magnitude of the perturbation. For the case
of $\alpha < 0.2$, this is below 10 $\%$, in other cases second-order
perturbations may be required.

For a more complete representation, it will be highly demanded to include
eccentricity (in progress) as well as higher-order spin dynamics,
which have been found recently by Steinhoff et al. {\cite{SSH08, SHS08}}.
Another important fact is that the radiation reaction (RR) has been neglected for
this analysis. It will be a task to include the energy and angular momentum loss
due to RR, which is under investigation.

\section{Acknowledgments}
I am grateful to thank Gerhard Sch\"afer for encouragement and
carefully reading of the manuscript and Jan Steinhoff and Steven Hergt
for helpful discussions. Special thanks are for Johannes Hartung for his
crosschecks.
This work was funded in part by the Deutsche Forschungsgemeinschaft (DFG)
through SFB/TR7 ``Gravitationswellenastronomie'' and the DLR
(Deutsches Zentrum f\"ur Luft- und Raumfahrt) through ``LISA Germany''.

\begin{appendix}
\section{solution of the full EOM by Lie series}

For a consistency check, let us solve the problem perturbatively using the Lie
series formalism {\cite{GrLe_LIE}} and compare the results with the
computation in the previous section.
The idea is to associate a
linear differential operator $\cal D$ to a system of differential equations
and to apply this operator in an exponential series to the initial values.
Successive computing of the addends will give the perturbed special solution
to the required order.
Let us suppose $\cal D$ to have the explicit form
\begin{equation}
 \label{eq::defin_diffop_D}
{\cal D} =	 \alpha_1(\vek{x}) \frac{\partial}{\partial x_1}+ ... 
		+\alpha_n(\vek{x}) \frac{\partial}{\partial x_n}\,,
\end{equation}
where $n$ is the number of independent variables and $\vek{x}=\{x_i\}$ with $i=(1,...,n)$.
The $\alpha_i$ are functions of these variables. Then the operator ${\cal D}$, applied
to the variable $x_i$, will give
\begin{equation}
 {\cal D}x_i=\alpha_i(\vek{x})\,.
\end{equation}
Under certain assumptions (holomorphy of the $\alpha_i$), the series
\begin{equation}
 \label{eq::def_LIE_series}
e^{t\,{\cal D}} f(\vek{x})
\equiv \sum_{\nu=0}^{\infty} \frac{t^\nu {\cal D}^{\nu}}{\nu !} f(\vek{x})
 = f(\vek{x}) + t {\cal D} f(\vek{x}) + \frac{t^2}{2!}{\cal D}^2 f(\vek{x}) ...
\end{equation}
converges absolutely and uniformly for some $|t|<T$. Defining $\vek{X}$
to be, in components,
\begin{equation}
 X_i \equiv (e^{t\,{\cal D}}{x_i}) \text{~with~} X_i|_{t=0}=x_i\,,
\end{equation}
the following ``exchange relation''
\begin{equation}
 F(\vek{X}) \equiv F(e^{t\,{\cal D}}\vek{x})=e^{t\,{\cal D}} F(\vek{x})
\end{equation}
holds for the region of convergence. Computing the time derivative of
the elements $X_i$, one can use the latter relation for the operator
$\partial_t$ as the function $F$ and obtains
\begin{equation}
\label{eq::DiffEq_LIE}
 \frac{d X_i}{d t}=e^{t {\cal D}}[{\cal D} x_i]=e^{t {\cal D}}[\alpha_i(\vek{x})]=\alpha_i(\vek{X})\,.
\end{equation}
This shows that the $X_i$ are solutions to Eq.. (\ref{eq::DiffEq_LIE}) in
the region of convergence for the time $t$.

The next step is to split the operator ${\cal D}$ into one part ${\cal D}_1$,
of which the solutions are exactly known, and another part ${\cal D}_2$ perturbing
this system of differential equations, both supposed to be holomorpic functions
in the same surrounding of the point $\vek{x}=\{x_1,...x_n\}$.
Let us define the solution to the operators ${\cal D}_1$ and ${\cal D}_2$ as
\begin{eqnarray}
 \label{eq::DE_ex+pert}
\vek{X}^{(0)}(t, \vek{x})&\equiv& e^{t\,{\cal D}_1}\vek{x}					\,,\\
 \label{eq::LIE_doube_series}
\vek{X}(t, \vek{x})&\equiv& e^{t\,{\cal D}} \vek{x}=e^{t\,({\cal D}_1 + {\cal D}_2)} \vek{x}	\,.
\end{eqnarray}
Inside the region of convergence, the series for $\vek{X}$ can be
resummed arbitrarily and cast into another more useful form
\begin{align}
\vek{X}(t, \vek{x})
		=& \vek{X^{(0)}}(t, \vek{x})
\nonumber \\ &
			+\sum_{\nu=0}^{\infty}\int_{0}^{t}
			d \tau
			 \frac{(t-\tau)^\nu}{\nu !}
			 ({\cal D}_2 {\cal D}^{\nu} \vek{x})|_{\vek{x}=\vek{X^{(0)}} }\,.
\end{align}
The series runs over the label $\nu$ and the operator ${\cal D}_2 {\cal D}^\nu$ has
to be applied to $\vek{x}$ {\em first}, the unperturbed solution $\vek{g}(t, \vek{x})$
to be inserted {\em afterwards},
{see \cite{GrLe_LIE} for further information}. The operator ${\cal D}_2$
can be shifted before the summation, which itself can also be exchanged with the integration,
and what remains is
\begin{equation}
 \vek{X}(t, \vek{x})
		= \vek{X}^{(0)}(t, \vek{x})
		+ \int_{0}^{t}
			d \tau
			 {\cal D}_2 \vek{X}(t-\tau, \vek{x})|_{\vek{x}=\vek{X}^{(0)}(\tau, x)}\,.
\end{equation}
This is an integral relation which can be solved iteratively. To any order,
for example, the solution reads
\begin{align}
\label{eq::LIE_any_order}
 \vek{X}^{(1)} (t, \vek{x}) =&~ \vek{X}^{(0)}(t,\vek{x}) \nonumber \\
			     &~ +\int_{0}^{t}
[{{\cal D}_2} \vek{X}^{(0)} (t-\tau, \vek{x})]_{\vek{x}=\vek{X}^{(0)}(\tau, \vek{x})}] d\tau \,,	~ \nonumber \\
\vdots \nonumber
\nonumber \\
\vek{X}^{(\nu+1)} (t, \vek{x})
			=&~ \vek{X}^{(0)}(t,\vek{x})  \nonumber \\
			&~ +\int_{0}^{t}
[{{\cal D}_2} \vek{X}^{(\nu)} (t-\tau, \vek{x})]_{\vek{x}=\vek{X}^{(0)}(\tau, \vek{x})}] d\tau \,. \hrulefill
\end{align}
For convergence issues, we note that this expression converges at least where
the double series, Eq. (\ref{eq::LIE_doube_series}), converges absolutely \cite{GR_LIE_appl}.
For a satisfying application of this algorithm, the operator ${\cal D}_2$ hast to be small;
that means that the functions $\alpha_i^{(2)}(\vek{x})$ (the superscript 2 stands for the
association to the second operator) are smaller in their magnitude in comparison to the
coefficients $\alpha_i^{(1)}(\vek{x})$.

This algorithm applies excellently to the problem of a binary of arbitrarily
configurated spins with unequal mass distribution, slightly deviating from
the exact equal-mass case. The latter is already
solved in \cite{KG05}, and what remains is to include perturbations.
We will, for the time being, resort to 
the first order of the approximation scheme
(\ref{eq::LIE_any_order})
to give a representative computation. Of course, the results will not sufficiently reflect
the physics of the system after a long elapsed time and has to be expanded for further investigations.

\noindent
To first-order in spin-orbit interactions, the motion of the spinning binary
can be split into the equal mass spin-orbit evolution
completed by the remainder built from
the difference in the masses. Let us choose $\vek{X}=\{\Theta, \Upsilon, \phi_s \}$ as the
functions to be evolved, then the EOM for $\vek{X}$, after the split, symbolically read
\begin{subequations}
\label{eq::EOM_SO_split}
	\begin{align}
	\dot \Theta	&= \chi_1\, {\cal T}_1(\Theta, \phi_s) + \chi_2\, {\cal T}_2(\Theta, \phi_s) \,, \\
	\dot \Upsilon	&= \chi_1\, {\cal U}_1(\Theta, \phi_s) + \chi_2\, {\cal U}_2(\Theta, \phi_s) \,, \\
	\dot \phi_s	&= \chi_1\, {\cal P}_1(\Theta, \phi_s) + \chi_2\, {\cal P}_2(\Theta, \phi_s) \,.
	\end{align}
\end{subequations}
The operators ${\cal D}_1$ and ${\cal D}_2$, therefore, read
\begin{eqnarray}
 {\cal D}_1 &\equiv \chi_1 \, \left ({\cal T}_1 \partial_{\Theta} +  {\cal U}_1 \partial_{\Upsilon} +  {\cal P}_1 \partial_{\phi_s} \right) \,, \\
 {\cal D}_2 &\equiv \chi_2 \, \left ({\cal T}_2 \partial_{\Theta} +  {\cal U}_2 \partial_{\Upsilon} +  {\cal P}_2 \partial_{\phi_s} \right) \,.
\end{eqnarray}
For the full motion, Eqs. (\ref{eq::EOM_SO_split}), then
$\vek{X}(t)$ is given by the Lie series
\begin{equation}
 \vek{X}(t)=e^{t\,({\cal D}_1 + {\cal D}_2)} \vek{X}(t=0) = e^{t\,({\cal D}_1 + {\cal D}_2)} \vek{x} \,.
\end{equation}
The relation for the perturbative functions can be computed using
the unperturbed one, associated with the equal-mass case. The generic angles therein,
$\vek{X^{(0)}}~=~\{ \Theta^{(0)},\Upsilon^{(0)}, \phi_s^{(0)} \}$, read
\begin{eqnarray}
 \Upsilon^{(0)}(t)		&=& \Omega_\Upsilon \, t \, + \Upsilon_{0}\,, \\
 \phi_s^{(0)}(t)		&=& \Omega_{\phi_s} \, t \, + {\phi_s}_{0} \,, \\
 \Theta^{(0)}(t)		&=& \Theta_{0}\,,
\end{eqnarray}
with {\em constant} angular velocities, given by Eqs. (\ref{eq::ang_vel}).
The first order solutions formally read
\begin{widetext}
\begin{subequations}
\label{eq::perturbing_LS1S2}
\begin{align}
 {\Theta}^{(1)}(t)- {\Theta^{(0)}}(t)	&= 
\int_0^t \left\{ {\cal D}_2 \,  {\Theta^{(0)}}(t-\tau, \vek{x})	\right \}_
{\vek{x}=\vek{X}^{(0)}(\tau, \vek{x})}	d \tau	\,, \\
{\Upsilon}^{(1)}(t)- {\Upsilon^{(0)}}(t)	&=
\int_0^t \left\{ {\cal D}_2 \,  {\Upsilon^{(0)}}(t-\tau,\vek{x})\right \}_
{\vek{x}=\vek{X}^{(0)}(\tau, \vek{x})}	d \tau	\,, \\
{\phi_s}^{(1)}(t)- {\phi_s^{(0)}}(t)	&=
\int_0^t \left\{ {\cal D}_2 \,  {\phi_s^{(0)}}(t-\tau,\vek{x})	\right \}_
{\vek{x}=\vek{X}^{(0)}(\tau, \vek{x})} d \tau	 \,. ~~~~~~~~~
\end{align}
\end{subequations}
All perturbing functions, computed by Eqs. (\ref{eq::perturbing_LS1S2}), are in complete agreement with the ones
in section \ref{sec::pert_sol}.
 \vspace{1cm}

\end{widetext}
\end{appendix}

\end{document}